\documentclass[twocolumn,showpacs,preprintnumbers,amsmath,amssymb,epsfig,widetext]{revtex4}

\usepackage{graphicx}% Include figure files
\usepackage{dcolumn}% Align table columns on decimal point
\usepackage{bm}% bold math
\usepackage{epsfig}
\usepackage{color}

\def\fun#1#2{\lower3.6pt\vbox{\baselineskip0pt\lineskip.9pt
  \ialign{$\mathsurround=0pt#1\hfil##\hfil$\crcr#2\crcr\sim\crcr}}}
\def\simgt{\mathrel{\lower0.6ex\hbox{$\buildrel {\textstyle >}
 \over {\scriptstyle \sim}$}}}
\def\simlt{\mathrel{\lower0.6ex\hbox{$\buildrel {\textstyle <}
 \over {\scriptstyle \sim}$}}}

\input epsf

\newcommand{\mnras}{MNRAS}
\newcommand{\apjl}{ApJL}
\newcommand{\aj}{AJL}

\def\be{\begin{equation}}
\def\ee{\end{equation}}
\def\ba{\begin{eqnarray}}
\def\ea{\end{eqnarray}}

\def\nn{\nonumber}

\begin{document}

\preprint{}

\title{Statistical Determination of Bulk Flow Motions}

\author{$^1$Yong-Seon Song, $^{1,2}$Cristiano G. Sabiu, $^1$Robert C. Nichol and $^2$Christopher J. Miller}
\email{yong-seon.song@port.ac.uk}
\affiliation{$^1$Institute of Cosmology $\&$ Gravitation, 
University of Portsmouth, Portsmouth, PO1 3FX, UK\\
$^2$Department of Physics \& Astronomy, University College London,
Gower Street, London, U.K \\
$^3$Cerro-Tololo Inter-American Observatory, National Optical 
Astronomy Observatory, 950 North Cherry Ave., Tucson, AZ 
85719, USA}

\date{\today}

\begin{abstract}
We present here a new parameterization for the bulk motions of galaxies and clusters (in the linear regime) that can be measured statistically from the shape and amplitude of the two--dimensional two--point correlation function. We further propose the one--dimensional velocity dispersion ($v_p$) of the bulk flow as a  complementary measure of redshift--space distortions, which is model--independent and not dependent on the normalisation method. As a demonstration, we have applied our new methodology to the C4 cluster catalogue constructed from Data Release Three (DR3) of the Sloan Digital Sky Survey. We find $v_p=270^{+433}\,$km/s (also consistent with $v_p=0$) for this cluster sample (at $\bar{z}=0.1$), which is in agreement with that predicted for a WMAP5--normalised $\Lambda$CDM model (i.e., $v_p(\Lambda{\rm CDM})=203\,$km/s). This measurement does not lend support to recent claims of excessive bulk motions ($\simeq1000$ km/s) which appear in conflict with $\Lambda$CDM, although our large statistical error cannot rule them out.  From the measured coherent evolution of $v_p$, we develop a technique to re-construct the perturbed potential, as well as estimating the unbiased matter density fluctuations and scale--independent bias.
\end{abstract}

\pacs{draft}


\maketitle

\section{Introduction}
A decade ago, astronomers discovered the expansion of the Universe was accelerating via the cosmological dimming of distant supernovae~\cite{Perlmutter:1998np,Riess:1998cb}. Since then,
the combination of numerous, and diverse, experiments
has helped to establish the Cosmological Constant (specifically a $\Lambda$CDM model) 
as the leading candidate to explain this cosmic acceleration. However, with no 
theoretical motivation to explain the required low energy vacuum of the $\Lambda$CDM model, 
there is no reason to preclude alternative models, especially those based upon the 
possible violation of fundamental physics which have yet to be proven on 
cosmological scales~\cite{dvali00,carroll05}. 

In addition to using geometrical probes like Supernovae to constrain the cosmic acceleration, tests based on the formation of structures in the Universe also provide a method for validating our cosmological models.
In particular, we can investigate the consistency 
between the geometrical expansion history of the Universe and the evolution of local density inhomogeneities to help reveal a deeper understanding of the nature of the cosmic acceleration~\cite{Song:2005gm,ishak05,knox05,linder05,Jain:2007yk}. 

In general, there are three observables that can be used to quantify
structure formation in the Universe, namely geometrical perturbations,
energy--momentum fluctuations and peculiar velocities, all of which
will be measured to high precision via future experiments like DES, LSST, JDEM and
Euclid (see details of these experiments in the recent FoMSWG report~\cite{Albrecht:2009ct}). In more detail, such weak lensing experiments
measure the integrated geometrical effect on light as its trajectory
is bent by the gravitational potential. Likewise, galaxies (and clusters of galaxies) measure the correlations amongst large--scale local inhomogeneities, while the observed distortions in these correlations (in redshift--space) can be used to extract information about peculiar velocities~\cite{Wang:2007ht,Guzzo:2008ac,2008arXiv0807.0810S,White:2008jy}. 
In this paper, we explore the cosmological constraints on the physics of cosmic acceleration using peculiar velocities, as it is one of the key quantities required for a consistency test of General Relativity~\cite{Song:2008vm,Song:2009zz}. 

Early observational studies of the peculiar velocity field, or ``bulk
flows", have produced for many years discrepant results~\cite{2007MNRAS.375..691S},
primarily due to small sample sizes and the heterogeneous selection of
galaxies. However, a recent re-analysis of these earlier surveys~\cite{Watkins:2008hf} has 
now provided a consistent observational picture from these data and finds significant evidence for a larger than expected bulk motion. This is consistent with new measurements of the bulk motion of clusters of galaxies using a completely different methodology\cite{Kashlinsky:2008ut,Kashlinsky:2009dw}, which leads to the intriguing situation that all these measurements appear to be significantly greater in amplitude, and scale, than expected in a concordance, WMAP5--normalised $\Lambda$CDM cosmological model. Such discrepancies with $\Lambda$CDM may give support to exotic cosmological models like modified gravity~\cite{Afshordi:2008rd}.

Given the importance of these large--scale bulk flow measurements, we propose here an alternative methodology to help check these recent claims of anomalously high peculiar velocities which are inconsistent with the standard $\Lambda$CDM cosmology. We start by outlining a statistical determination of  bulk flow motions using redshift--space distortions in large-scale galaxy or cluster surveys. Such redshift-space distortions are easily seen in the two--dimensional correlation function ($\xi_s(\sigma,\pi)$), which is the decomposition of the correlation function into two vectors; one parallel ($\pi$) to the line--of--sight and the other perpendicular ($\sigma$) to the line--of--sight. On small scales, any incoherent velocities of galaxies within a single dark matter halo (or cluster) will just add to the cosmological Hubble flow thus causing the famous ``Fingers-of-God" (FoG) effect which stretches the 2-D correlation function preferentially in line-of-sight ($\pi$) direction. These distortions depend on the inner dynamics and structure of halos and therefore, any cosmological 
information is difficult to distinguish from the halo properties. However, on large scales (outside individual dark matter halos), the peculiar velocities become coherent and follow the linear motion of the matter thus providing crucial information on the formation of large-scale structure~\cite{Kaiser:1987qv}.

In this paper, we compare predictions for $\xi_s(\pi,\sigma)$ to observations based on the C4 cluster catalogue~\cite{2005AJ....130..968M} from the Sloan Digital Sky Survey (SDSS)~\cite{York:2000gk}.
Using the Kaiser formulation~\cite{Kaiser:1987qv}, 
a theoretical model for $\xi_s(\pi,\sigma)$ is fit to the measured 2-D correlation function in configuration--space, with the $\xi_s$ parameterised by a shape dependent part 
and a coherent evolution component. We also propose that the 1-D linear velocity dispersion ($v_p$) is a interesting quantity to
report when measuring redshift--space distortions, and complementary to traditional
quantities like $\beta$, $f$ or $f\sigma_8$ discussed recently~\cite{Wang:2007ht,Guzzo:2008ac,2008arXiv0807.0810S}, as it is independent of both bias
and the normalisation method. Therefore, the measured $v_p$ provides an unbiased tracer of the evolution of structure formation. 

\section{Statistical determination of peculiar velocity}

The redshift--space two-point correlation function of mass tracers ($\xi_s(\sigma,\pi)$) is an anisotropic function~\cite{Kaiser:1987qv}. On small scales, it is elongated in the $\pi$-direction by the ``Fingers-of-God" effect, while on large scales, the gravitational
infall into overdense regions preferentially compresses the correlation function in the $\sigma$ direction. Therefore, peculiar velocities can be statistically measured
by analyzing the observed anisotropic pattern of $\xi_s(\sigma,\pi)$ in both
the linear and non-linear regimes.

$\xi_s(\sigma,\pi)$ is derived from the
convolution of $\xi(r)$ with a probability distribution
function of peculiar velocities along the line of sight, which is usually called
the streaming model~\cite{Davis:1982gc}. Even with the simplest form of a 
Gaussian probability distribution, the streaming model
describes the suppression effect on $\xi_s(\sigma,\pi)$ on small scales. 

In the
linear regime, the density fluctuations and peculiar velocity are
coherently evolved through the continuity equation, which is known as the 
Kaiser limit. Thus the known
correlation function of $\xi(r)$ from the linear perturbation theory
developed by gravitational instability is uniquely transformed into
$\xi_s(\sigma,\pi)$~\cite{1989MNRAS.236..851L,1990MNRAS.242..428M,1991MNRAS.251..128L,1992ApJ...385L...5H,1994MNRAS.266..219F}.

The large scale limit of the streaming model is consistent with the
Kaiser limit~\cite{Kaiser:1987qv}, when both the density and peculiar velocity
fields are treated as statistical quantities~\cite{1995ApJ...448..494F}. This consistency
test was developed further to show that, even in the Kaiser limit, the
description of $\xi_s(\sigma,\pi)$ in linear theory can be modified
due to the correlation between the "squashing" (in the $\sigma$ direction)  and dispersion
effects (in the $\pi$ direction)~\cite{Scoccimarro:2004tg}. With the assumption of a Gaussian pair-wise
velocity distribution function, the dispersion effect smears into the 
Kaiser limit description of $\xi_s(\sigma,\pi)$ at around the percent
level which for our present work can be ignored. Thus we adopt the Kaiser limit for the description of
$\xi_s(\sigma,\pi)$ in linear regime while considering dispersion
effect as a systematic uncertainty.
We introduce below a new parameterisation of $\xi_s(\sigma,\pi)$ in
terms of the cosmological parameters and construct a method to
measure the mean velocity dispersion $v_p$ in a model independent
way. 

\subsection{Model independent parameterisation of power spectra}

\begin{figure}[t]
 \begin{center}
 \epsfysize=3.0truein
 \epsfxsize=3.0truein
   \epsffile{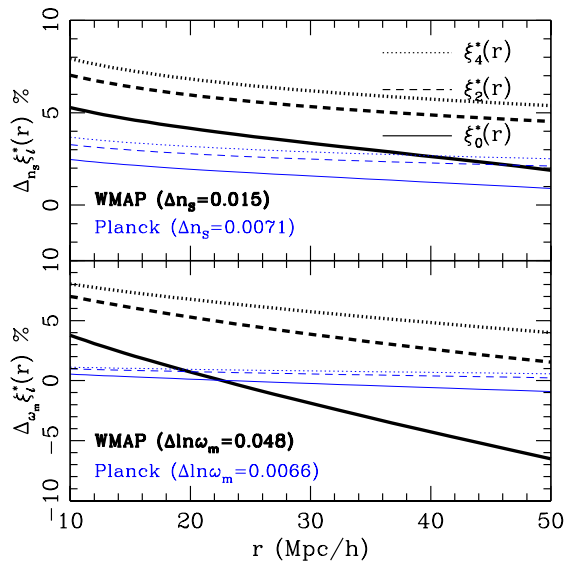}
   \caption{\label{fig:xi_diff}\footnotesize $\Delta\xi_l^*(r)$ for various CMB
     experiments priors. In the upper panel, we show the change in
     $\xi_l^*(r)$ for variations in $\omega_m$. In the
     lower panel, we show the change in $\xi_l^*(r)$ as a function of  
     $n_S$. The thick black curves are based on WMAP
      priors, while the thin blue curves are for Planck prior.  
      On the y-axis, we focus on the range of scales probed by recent and planned reshift--space distortion measurements.}
\end{center}
\end{figure}

The discovery of cosmic acceleration has prompted rapid progress in theoretical cosmological research and prompted many authors to propose modification to the law of gravity beyond our solar system. For example, some theories based upon General 
Relativity can be modified by screening (or anti-screening) the mass of gravitationally 
bound objects~\cite{dvali00}, while others include a non-trivial dark energy component (e.g. interacting dark energy~\cite{Amendola:1999er,Koyama:2009gd}, or clumping dark energy~\cite{Kunz:2006ca}) thus breaking the dynamical relations between density fluctuations and peculiar velocity in the simplest dark energy models. These theoretical ideas motivate us to express various power spectra of the density field in a more convenient way to test such theories. 

We assume a standard cosmology model for epochs earlier than the last
  scattering surface, and that the coherent
  evolution of structure formation from the last scattering surface to
  the present day is undetermined due to new physics relevant to the
  cosmic acceleration. Thus we divide the history of structure formation into two regimes; epochs before matter-radiation equality ($a_{eq}$) and a later epoch of coherent evolution
  of unknown effect on structure formation from new physics.We can then express various power spectra of the density field splits into these two epochs, with the shape-dependent part determined by knowledge of our standard cosmology, and the coherent evolution part only affected by new physics. Mathematically, this is written as, 
\ba
P_{\Phi\Phi}(k,a)&=&D_{\Phi}(k)g_{\Phi}^2(a),\nn\\
P_{bb}(k,a)&=&D_{m}(k)g_{b}^2(a),\nn\\
P_{\Theta_m\Theta_m}(k,a)&=&D_m(k)g_{\Theta_m}^2(a)\,,
\ea
where $\Phi$ denotes the curvature perturbation in the Newtonian gauge, 
\begin{equation}
 ds^2=-(1+2\Psi)dt^2+a^2(1+2\Phi)dx^2\,,
\end{equation}
%and $b$ denotes density inhomogenous field of tracer (e.g. cluster or galaxy) 
and $\Theta_m$ denotes the map of 
the re-scaled divergence of
peculiar velocity $\theta_m$ as $\Theta_m=\theta_m/aH$.
These power spectra are then partitioned into a scale--dependent
part ($D_{\Phi}(k),D_{m}(k)$) and a scale-independent (coherent evolution) component ($g_{\Phi}$, $g_{b}$, $g_{\Theta_m}$). We define here $g_{b}=b\,g_{\delta_m}$ where $b$ is the standard linear bias parameter between galaxy (or cluster) tracers and the underlying dark matter density.

The shape of the 
power spectra is determined before the epoch of matter--radiation equality. Under the
paradigm of inflationary theory, initial fluctuations are stretched outside the
horizon at different epochs which generates the tilt in the power
spectrum. The predicted initial tilting is parameterised as a spectral
index ($n_S$) which is just the shape dependence due to the initial
condition. When the initial fluctuations reach the coherent evolution epoch after matter-radiation
equality, they experience a scale-dependent shift from the moment they re-enter the 
horizon to the equality epoch. Gravitational instability is governed by the interplay 
between radiative pressure resistance and gravitational infall. The different
duration of modes during this period results in a secondary shape
dependence on the power spectrum. This shape dependence is determined by
the ratio between matter and radiation energy densities and sets the
location of the matter-radiation equality in the time coordinate. As the 
radiation energy density is precisely measured by the CMB blackbody
spectrum, these secondary shape dependences are parameterised by
the matter energy density $\omega_m=\Omega_mh^2$. Both of these parameters are now well--determined by CMB experiments. 

The shape factor of the perturbed metric power spectra $D_{\Phi}(k)$ is
defined as
\be
D_{\Phi}(k)=\frac{2\pi^2}{k^3}\frac{9}{25}\Delta^2_{\zeta_0}(k)T^2_{\Phi}(k)
\ee
which is a dimensionless metric power spectra at $a_{eq}$, where
$\Delta^2_{\zeta_0}(k)$ is the initial fluctuations in the comoving gauge and 
$T_{\Phi}(k)$ is transfer function normalized at
$T_{\Phi}(k\rightarrow 0)=1$. 
The primordial shape
$\Delta^2_{\zeta_0}(k)$ depends on $n_S$, as
$\Delta^2_{\zeta_0}(k)=A^2_S(k/k_p)^{n_S-1}$, where
$A^2_S$ is the amplitude of the initial comoving fluctuations at the pivot scale, 
$k_p=0.002 {\rm Mpc}^{-1}$. The intermediate shape factor
$T_{\Phi}(k)$ depends on $\omega_m$. The shape factor for matter
fluctuations and peculiar velocities $D_m(k)$ are given by the
conversion from $D_{\Phi}(k)$ of,
\be
D_m(k)\equiv\frac{4}{9}\frac{k^4}{H_*^4\omega_m^2}D_{\Phi}(k),
\ee
where $H_*=1/2997\,{\rm Mpc}^{-1}$. 

Unlike the shape part, the coherent evolution component, $g_{\Phi}$, $g_{b}$ and
$g_{\Theta_m}$ are not generally parameterized by known standard
cosmological parameters.
We thus normalize these growth factors at $a_{eq}$ such that,
\ba
g_{\Phi}(a_{eq})&=&1,\nn\\
g_{\delta_m}(a_{eq})&=&a_{eq}g_{\Phi}(a_{eq}), \nn\\
g_{\Theta_m}(a_{eq})&=&-\frac{dg_{\delta_m}(a_{eq})}{d\ln a}\,.
\ea
Instead of determining growth factors using cosmological parameters, 
we measure these directly in a model-independent way
at the given redshift without referencing to any specific cosmic
acceleration model and with the minimal assumption of coherent evolution of
modes after $a_{eq}$. Considering the uncertainty in the determination of
$A_S^2$ from the CMB anisotropy, which is degenerate with the optical
depth of re-ionization, we combine both $A_S^2$ and $g_X$ (where $X$ denotes
each component of $\Phi$, $b$ and $\Theta_m$) with proper scaling for
convenience as $g_X^*=g_XA_S/A_S^{*}$. 
Throughout this paper, we
use $A_S^{*\,2}=2.41\times 10^{-9}$ for the mean $A_S^2$ value from the WMAP5 results.
Our result on measuring the bulk flow motion is independent of our choice of
an arbitrary constant $A_S^{*\,2}$. 

\subsection{Correlation function in the configuration space}

\begin{figure}[t]
 \begin{center}
 \epsfysize=3.3truein
 \epsfxsize=3.3truein
   \epsffile{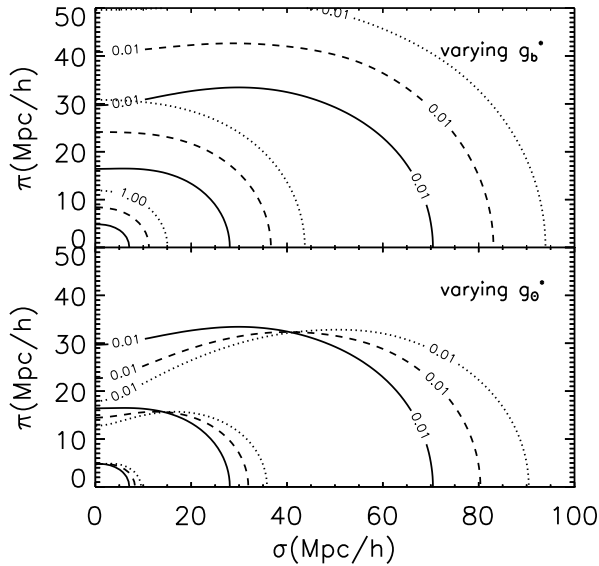}
   \caption{\label{fig:sigma_pi_theory}\footnotesize The  $\xi_s(\sigma,\pi)$~correlation function. We plot for three contours of  $\xi_s(\sigma,\pi)= 1, 0.1, 0.01$ (from the inner to outer contour). In the upper
   panel, the solid curves are for a $\Lambda$CDM cosmology, while the dash and dotted curves are for models with $g_b= 1.5$ and $2$ respectively. In the lower
   panel, the solid curves are for a $\Lambda$CDM cosmology, while the dash and dotted curves are for models with $g_{\Theta_m}=1.5$ and $2$ respectively.} 
\end{center}
\end{figure}

In the linear regime of the
standard gravitational instability theory, the Kaiser effect (the observed squeezing of $\xi_s(\sigma,\pi)$ due to coherent infall around large--scale structures) can be written in configuration space  as,
\ba\label{eq:xis}
\xi_s(\sigma,\pi)(a)&=&
\left(g_b^{*\,2}+\frac{1}{3}g_b^*g_{\Theta_m}^*+\frac{1}{5}g_{\Theta_m}^{*\,2}\right)\xi_0^*(r){\cal
  P}_0(\mu)\nn\\
&-&\left(\frac{4}{3}g_b^*g_{\Theta_m}^*+\frac{4}{7}g_{\Theta_m}^{*\,2}\right)\xi_2^*(r){\cal
  P}_2(\mu)\nn\\
&+&\frac{8}{35}g_{\Theta_m}^{*\,2}\xi_4^*(r){\cal P}_4(\mu),
\ea
where ${\cal P}_l(\mu)$ is the Legendre polynomial and the spherical
harmonic moment $\xi_l^*(r)$ is given by,
\be
\xi_l^*(r)=\int\frac{k^2dk}{2\pi^2}D_m^*(k)j_l(kr),
\ee
where $j_l$ is a spherical Bessel function and $*$ denotes scaling of the
shape factor with $A_S^{*\,2}$.

As discussed above, we ignore the effect on $\xi_s(\sigma,\pi)$ of the small-scale velocity dispersions within a single dark matter halo~\cite{Scoccimarro:2004tg} as the effect is only a few percent, and split Eqn.~\ref{eq:xis} into a shape--dependent part ($\xi_l^*(r)$), which is determined by the 
cosmological parameters ($n_S$ and $\omega_m$), and a coherent
evolution component, which is parameterised by $g_{b\,i}^*$, $g_{\Theta_m\,i}^*$
at the targeted redshift $z_i$. The shape part is therefore almost completely determined
by CMB priors, while the coherent
evolution of structure formation can be determined from fitting $\xi_s(\sigma,\pi)$, in redshift--space, as a function of redshift.

In Figure ~\ref{fig:xi_diff}, we present the effect of CMB priors on the value of $\xi_l^*(r)$. In the top panel of Fig.~\ref{fig:xi_diff}, we provide the expected variation in $\xi_l^*(r)$ from varying $\omega_m$. We see that varying $a_{eq}$ causes greater tilting in the shape of $\xi_l^*(r)$, since larger scale modes can come into the
horizon earlier. 
%Thus a stiffer shape of $\xi_l^*(r)$ is
%expected with higher values of $\omega_m$ as shown in
%Fig.~\ref{fig:xi_diff}. 
In addition to this contribution, the overall amplitude of $\xi_l^*(r)$
depends on $\omega_m$ by a weighted transformation between $D_{\Phi}(k)$
and $D_m(k)$. Considering the marginalisation over CMB priors, we expect a discrepancy
of $\simeq5\%$ with WMAP5 measurements, and just a few percent effect with the projected Planck priors.

In the bottom panel of Fig.~\ref{fig:xi_diff}, the dependence of $\xi_l^*(r)$ on $n_S$ is given for both WMAP5 and Planck
priors. The overall shifting can be re-scaled by adjusting the pivot
point to the effective median scale of the survey. With the measured WMAP5 prior of
$\Delta n_S=0.015$~\cite{Komatsu:2008hk}, we expect variations of a few percent on the shape, while for  an estimated Planck prior of $\Delta n_S=0.0071$~\cite{Kaplinghat:2003bh}, we expect $\xi_l^*(r)$ to be nearly invariant to $n_S$. The shape of $\xi_l^*(r)$
is affected maximally during the intermediate epoch, from horizon crossing to 
the matter-radiation epoch. The decay rate of the inhomogeneities differs by the ratio between
matter and radiation energy densities.

Once CMB constraints are placed on the shape part of Eqn. 6,  the coherent history of structure
formation is obtained from the anisotropic moment of
$\xi_s(\sigma,\pi)$. Even though both $g_{b\,i}^*$ and
$g_{\Theta_m\,i}^*$ weight the evolution sector simultaneously, their
contribution to $\xi_s(\sigma,\pi)$ are different, which
enables us to discriminate $g_{\Theta_m\,i}^*$ from $g_{b\,i}^*$. In the 
monopole moment, $g_{b\,i}^*$ is the dominant component since
$g_{b\,i}^*>g_{\Theta_m\,i}^*$ unless their is an excessive bulk flow. 
Thus the variation of $g_{b\,i}^*$ generates a near isotropic
amplification as illustrated in the top panel of
Figure~\ref{fig:sigma_pi_theory}. In the quadrupole moment, the
cross-correlation between $\delta_m$ and $\Theta_m$ is leading
order. The reversed sign of the quadrupole moment results in the squashing
effect, and it is sensitive to the variation of $g_{\Theta_m\,i}^*$ as
the cross-correlation is the leading order. In the bottom panel of
Figure~\ref{fig:sigma_pi_theory} the variation of $g_{\Theta_m\,i}^*$
mainly contributes to the anisotropic moment. It is this signal which allows 
both $g_{b\,i}^*$ and $g_{\Theta_m\,i}^*$ to be probed separately using the 
anisotropic structure of $\xi_s(\sigma,\pi)$. 
The contribution from the term having peculiar velocity
autocorrelation is not significant if excessive bulk flows are
excluded. 

\subsection{Implication for cosmology from measuring $g_{b\,i}^*$ and
  $g_{\Theta_m\,i}^*$}

A measurement of $g_{\Theta_m\,i}^*$ is equivalent to the quantity $f\sigma_8^{\rm mass}$~\cite{White:2008jy} and therefore,
an excellent test of dark energy models (where $f$ is the logarithmic
derivative of the linear growth rate and $\sigma_8^{\rm mass}$ is the
root-mean-square mass fluctuation in spheres with radius
$8h^{-1}$Mpc).
While cosmological test of $g_{\Theta_m\,i}^*$ are free from bias, which
is notoriously difficult to measure accurately in a model--independent way,
the reported value of $g_{\Theta_m\,i}^*$ does depend on the
normalization which is also poorly constrained (i.e., primordial amplitude) or model--dependent 
(i.e., $\sigma_8^{\rm mass}$).

\begin{figure}[t]
 \begin{center}
 \epsfysize=3.truein
 \epsfxsize=3.truein
 \epsffile{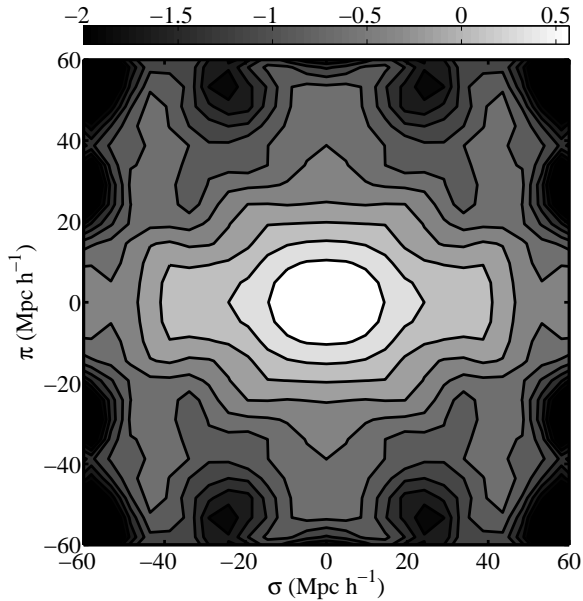}
   \caption{\label{fig:sp}\footnotesize The 2-D two--point correlation function ($\xi_s(\sigma,\pi)$) for the SDSS DR3 C4 cluster survey with a median redshift of ${\bar z}$=0.1. The contours have been slightly smoothed.}
\end{center}
\end{figure}

Thus, we introduce a more convenient parameterisation of peculiar
velocity which is independent of these normalization issues.
The measured $g_{\Theta_m\,i}^*$ (in the redshift bin $i$) which can be translated into
the one-dimensional (1-D) velocity dispersion in that redshift bin ($v_p^i$) 
by,
\be\label{eq:vp}
v_p^{i\,2}=g_{\Theta_m\,i}^{*\,2}\frac{H^2}{3}\int^{\infty}_0\frac{dk}{k}D_{m}(k)dk.
\ee
In this formula, there is a degeneracy between $g_{b\,i}^*$ and $g_{\Theta_m\,i}^*$ which cannot be solely broken by fitting $\xi_s(\sigma,\pi)$; instead we simultaneously fit for $ v_p^{i}$ and $b^i$ from the data and then marginalize over the bias to obtain $v_p$ (independent of $b$) in that redshift bin. Therefore, if our statistical determination of the history of $v_p^i$ can be combined with an
independent measurement of bias, then $v_p^i$ can be determined
precisely. The scaled parameter $g_{\Theta_m\,i}^*$ depends on all
shift factors; the primordial amplitude or the enhancement
of $D_m$ due to varying $\omega_m$, as well as later time $\Theta_m$
evolution. But the estimation of $v_p$ from $g_{\Theta_m\,i}^*$ is
independent of  the uncertainty in the overall shifting.

If the evolution of $g_{\Theta_m}^*$ is measured, it can be used to
reconstruct other coherent growth factors. The coherent growth factor
of $\Phi$ can be given using the Euler equation,
\be\label{eq:Euler}
g_{\Phi}^*=\frac{2}{3}\frac{aH}{H_*^2\omega_m}
\left(g_{\theta_m}^*+\frac{dg_{\theta_m}^*}{d\ln a}\right),
\ee
where no anisotropy condition is used. If the Poisson equation is
validated then the re-constructed $g_{\Phi}^*$ can be used to derive
$g_{\delta_m}^*$ using the relation
$g_{\delta_m}^*=ag_{\Phi}^*$. Then this estimated matter fluctuation
evolution can be used to determine bias from the measured
$g_{b\,i}^*$ as $b=g_{b\,i}^*/g_{\delta_m}^*$.

\section{Redshift--space distortions from clusters of galaxies}

As a demonstration of the parameterization discussed above, we present here a measurement of the redshift--space 2-D two--point correlation function ($\xi_s(\sigma,\pi)$) for clusters of galaxies selected from the SDSS.
We use an updated version of the C4 cluster catalogue \cite{2005AJ....130..968M} based on Data Release 3 (DR3; \cite{2007ApJS..172..634A}) of the SDSS. Briefly, the C4 
catalogue identifies clusters in a seven-dimensional galaxy position 
and colour space (righ ascension, declination, redshift, $u-g, g-r, r-i, i-z$) using the SDSS Main Galaxy 
spectroscopic sample. This method greatly 
reduces the twin problems of projection effects and redshift space distortions 
in identifying physically--bound galaxy groups. This catalogue is composed of $\sim2000$~clusters in the redshift range $0.02<z<0.15$.

The estimation of the correlation function relies crucially on our ability to 
compare the clustering of the data to that of a random field. Thus any 
artificial structures in the data must be considered when constructing the 
random catalogue. These problems include incompleteness, such 
as the angular mask (e.g. survey boundaries, bright stars and dust extinction 
in our own galaxy), and the radial distribution where at large distances, the 
mean space density decreases as we approach the magnitude limit of the survey.
We have constructed random samples which takes these issues into 
account, i.e., the angular positions are randomly sampled from a sphere to lie within the DR3 mask, while the redshifts are 
obtained from a smooth spline fit to the real C4 redshift distribution (which removes true large scale structures). The random samples are then made to be 50 times denser than the real data to avoid Poisson noise. 

In Figure~\ref{fig:sp}, we show our estimation of the $\xi_s(\sigma,\pi)$ binned into 
with 6 configuration--space bins up to 60 Mpc (one bin per 10Mpc). Separations of less than 10 Mpc are removed to reduce the FoG effect. Error on $\xi_s(\sigma,\pi)$ were derived using the jackknife method~\cite{2002ApJ...579...48S}, which involves dividing the survey into $N$ sub-sections with equal area (and thus volume) and then computing the mean and variance of $\xi_s(\sigma,\pi)$ from these $N$ measurements of the correlation function with the 
$i^{th}$ region removed each time (where $i=1...N$).

\begin{figure}[t]
 \begin{center}
 \epsfysize=3.0truein
 \epsfxsize=3.0truein
   \epsffile{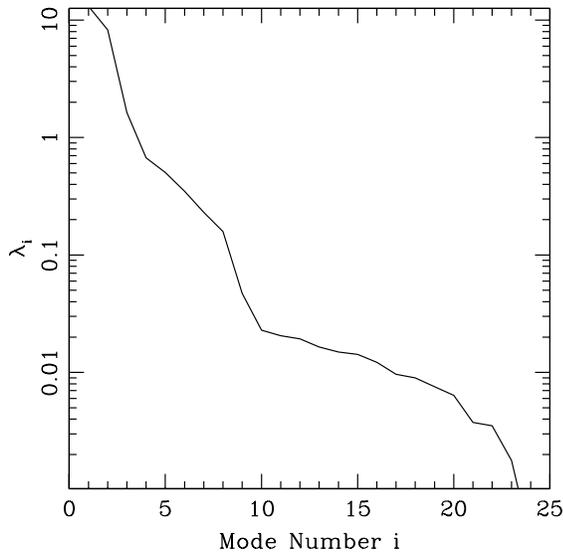}
   \caption{\label{fig:eigenmode} \footnotesize An ordered list of the Eigenvalues for our cluster
     covariance matrix.}
\end{center}
\end{figure}

In our analysis, we divided the whole C4 area into $N=30$ sub-subsections and determine the  variance from~\cite{1993stp..book.....L}, 
\begin{equation}
\sigma^2_\xi(r_i)=\frac{N_{jack}-1}{N_{jack}}\sum^{N_{jack}}_{k=1}[\xi_k(r_i)-\overline{\xi}(r_i)]^2,
\end{equation}
where $N_{jack}$ is the number of jackknife samples used and $r_i$ represents a single bin in the $\sigma-\pi$ configuration space. Then we compute
\begin{equation}
\overline{\xi}(r_i)=\frac{1}{N_{jack}}\sum^{N_{jack}}_{k=1}\xi_k(r_i),
\end{equation}
and the normalised covariance matrix is estimated
from~\cite{2002ApJ...579...48S}, 
\begin{equation}\label{eq:Cij}
C_{ij}=\frac{N_{jack}-1}{N_{jack}}\sum^{k=N_{jack}}_{k=1}\Delta^k_i\Delta^k_j,
\end{equation}
where,
\begin{equation}
\Delta^k_i=\frac{\xi_k(r_i)-\overline{\xi}(r_i)}{\sigma_{\xi}(r_i)}.
\end{equation}

Before we invert $C_{ij}$ in Eqn.~\ref{eq:Cij}, we note that the values of  $C_{ij}$ are estimated to limited resolution,
\begin{equation}
\Delta C_{ij}=\sqrt{\frac{2}{N_{jk}}}
\end{equation}
and therefore, if $N_{jk}$ is small, or there are degeneracies within  $C_{ij}$, the inversion will be affected. This problem can be eliminated by performing a Single Value Decomposition (SVD) of the matrix,
\begin{equation}
 C_{ij} =  U_{ik}^{\dag} D_{kl} V_{lj},
 \end{equation}
 where $U$ and $V$ are orthogonal matrices that span the range and
the null space of $C_{ij}$ and $D_{kl}=\lambda^2\delta_{kl}$, a diagonal
 matrix with singular values along the diagonal. In doing the SVD, we
 select the dominant modes to contribute to the $\chi^2$ by requiring
 that $\lambda^2 > \sqrt{2/N_{jk}}$.

In Figure~\ref{fig:eigenmode}, we rank the eigenvalues ($\lambda_i$) for
the increasing eigenmodes and see a ``kink" in the distribution which we
interpret as indicating a transition in the signal--to--noise of the eigenmodes, i.e., only the first ten modes contain most of the signal, while higher-ordered modes are dominated
by noise. We therefore remove eigenmodes beyond this kink (with $\lambda_i < 0.01$) where the eigenvalues start to flatten out.

\subsection{Statistical determination of large scale flow}

As discussed in the Introduction, there is recent evidence for excessive bulk flow motions compared to the WMAP5-normalised $\Lambda$CDM model  \cite{Watkins:2008hf} and therefore, it is important to confirm these results as it may indicate evidence for an alternative explanation for the observed cosmic acceleration  such as modified gravity. In this paper, we provide a first demonstration of our new
parameterization using clusters of galaxies from the SDSS. In detail,
we attempt to model the ``squashing" of the 2-D correlation function of the C4 cluster sample seen in Figure 5 using the formalism presented herein. We do however caution the reader that we expect the limited size of the DR3 sample to leads to large statistical errors, due to a significant
shot--noise contribution because of their low number density. However, future cluster and galaxy samples (e.g., LRGs) should provide stronger constraints and provide a more robust test of these high bulk flow measurements in the literature.

\begin{figure}[t]
 \begin{center}
 \epsfxsize=3.0truein
   \epsffile{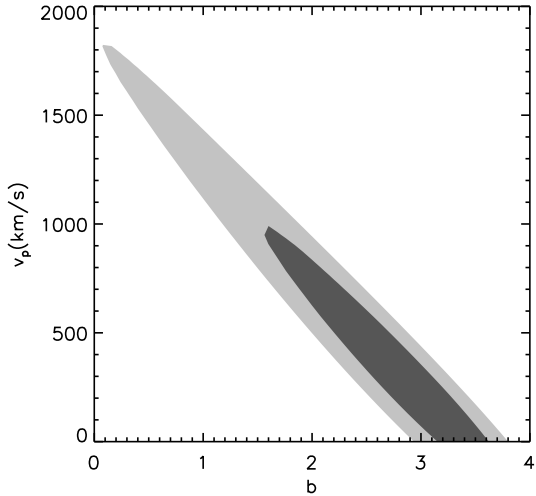}
   \caption{\label{fig:contour_fit} \footnotesize The 2-D contours between $b$ and
     $v_p$ with the DR3 cluster sample.}
\end{center}
\end{figure}

In Figure~\ref{fig:contour_fit}, we provide the best fit parameters $b$ and
$v_p$ for the C4 correlation function presented in Figure 3 and there is as expected a clear anti-correlation between these two
parameters because the anisotropic
amplitude is generated by cross-correlations in the density and
peculiar velocity fields. The best fit value from Figure 5 is $v_p=270^{+433}$ km/s (at the 1$\sigma$
level marginalised with other parameters including $b$) and is consistent with $v_p=0$. We do not quote the negative bound of the error on $v_p$ as it is below zero and thus has no physical meaning. Instead, we quote the upper bound on $v_p$ and note that our result is consistent with zero. Our measurement of $v_p$ is close to the predicted value of 203 km/s for a WMAP5--normalised $\Lambda$CDM model. 

We propose above that $v_p$ is a complementary parameter for reporting such peculiar velocity
measurements. The parameter $g_{\Theta}$, which is equivalent to
$f\sigma_8$, is not determined precisely without the prior information of
$A_S$. But when we report our measurement with $v_p$, there is no
uncertainty due to other cosmological parameters which are not
determined accurately, 
as it is equivalent to $g^*_{\Theta}$ determined statistically from
  redshift space distortion. 
The observed value
$v_p$ at a given redshift is not only independent of bias but
also independent of normalisation.

\subsection{Reconstruction of matter density field from $v_p$}

We convert $v_p$ measurement into $g_{\Theta}$ using $A_s$
from WMAP5 
($g_{\Theta}$: coherent growth factor of peculiar velocity,
and it is equivalent to $f\sigma_8$ in other parameterizations). With
the evolution of $g_{\Theta}$ known, dynamics of perturbations are
reconstructed to provide the history of $\Psi$ through the Euler
equation. In most theoretical models, the time variation of $v_p$ is
minimal at these low redshifts discussed here for the C4 sample ($z\simeq0.1$),
which allows us to ignore the time-derivative part in
Eq.~\ref{eq:Euler}. Therefore, it is straightforward to transform the coherent
evolution of $\Theta$ into the coherent evolution of $\Psi$. If we
assume no anisotropic stress, then it is easy to convert to the
coherent evolution of $\Phi$, $g_{\Phi}$.

We are able to determine matter density fluctuations through the 
Poisson equation. We calculate
the coherent growth of $\delta_m$, $g_{\delta}=0.7$, which is related
to $g_{\Phi}$ as $g_{\delta}=ag_{\Phi}$, if no modified Poisson
equation is assumed.
Finally, the estimated $g_{\delta}$ can be used to derive bias using
measured $g^*_b$. Through fitting to the redshift distortion effect,
we extract both $g^*_b$ and $g^*_{\Theta}$. The density fluctuation
evolution $g_{\delta}$ is estimated only from $g^*_{\Theta}$, and the
other measurement $g^*_b$ is not yet used. The combination of
the estimated $g_{\delta}$ and the measured $g^*_{\Theta}$ provides bias
from $b=g^*_b/g_{\delta}= 2.9\pm0.8$, which is fully consistent with our expectations for such massive clusters of galaxies in the C4 sample.

\section{Discussion}

We outline in this paper a new theoretical model for $\xi_s(\pi,\sigma)$, the 2-D two--point correlation function in configuration--space, which allows us to constrain the bulk flow motion of matter on large scales. We also propose that the 1-D linear velocity dispersion ($v_p$) is a interesting quantity to
report when measuring redshift--space distortions. We demonstrate this
method using C4 clusters from the SDSS and find a value for $v_p$ that
is consistent with a WMAP5--normalised $\Lambda$CDM cosmology (within
our large statistical errors).  Our observed value for these bulk flows
is marginally inconsistent
with other recent observations in the literature, which find an excess flow compared to a WMAP5--normalised $\Lambda$CDM model~\cite{Watkins:2008hf,Kashlinsky:2008ut,Kashlinsky:2009dw}. We do not discuss this further as we plan to revisit these measurements using larger datasets and different tracers of the density field.

As discussed in Section II-C, our measurement of $v_p$ is correlated with $g_b$, which is the combination of $b$ and $\delta_m$, since the observed anisotropic shape
of the 2-D correlation function in redshift--space is generated by a cross
correlation between the density field and peculiar
velocities. However, one of the important implications of our method is that we can measure $v_p$ without knowing how to decompose $g_b$, and thus without the uncertainty of determining $b$. 
%That said, it is premature to dismiss observed high cluster bulk flow motions of around 1000 km/s due to the anti-correlation between bias and $v_p$ (see Fig. 5). If we can measure the cluster
%bias in an alternative way, and it favors a lower bias, then $v_p$
%could rise and be determined to greater precision.

There are however some caveats to our analysis. We do not analyse our data in
Fourier space, but in configuration space. Small scales have been removed from our data analysis ($<10$Mpc),
to ensure that the FoG effect will not contaminate our results. Our methodology is insensitive to a 
possible shape dependence at large scales for any exotic reason; scale dependent
later time growth (e.g. f(R) gravity models~\cite{Song:2006ej}), or scale--dependent bias
at large scales. In follow-up studies, we will measure the redshift--space distortions in Fourier space to test the effect of small--scales on our results. In addition,
the formulation to derive $\xi_s$ used in this paper can be slightly
biased due to the dispersion effect studied
in~\cite{Scoccimarro:2004tg}. This effect is not parameterised properly here, but the reported level of uncertainty is approximately 5$\%$ which is much smaller than the statistical errors on our present measurements. Therefore, we dismiss this shift here but it is worth revisiting this issue in the future to know how to incorporate this effect in a new parameterisation. 

Finally, Song and Percival~\cite{2008arXiv0807.0810S} recently proposed a method 
to re-construct the structure formation observables from $\Theta$ measurements. 
Although it is not yet estimated precisely in that narrow range of measured values, we apply
their methodology in practice. From the observed coherent evolution of
$\Theta$ at $z=0.1$ from the DR3 C4 clusters, we re-construct $\Psi$,
$\Phi$ and $\delta_m$. We then find that bias can be derived from the
estimated $\delta_m$ and the measured $g_b^*$. Here, for the first
time, we estimate bias from peculiar velocity measurements only. The
estimated values are reasonable at
$b=g^*_b/g_{\delta}=2.9\pm 0.8$. It is not 
precise measurement yet, as the time variation is ignored, but 
we will revisit this in a following paper.

\section*{Acknowledgments}

The authors would like to thank Nick Kaiser, Kazuya Koyama and Will Percival for
helpful conversations,  
Prina Patel for useful suggestions on the presentation of this
article,
and the referee for helpful comments.
Y-SS, RCN and 
CGS are grateful for support from STFC .

%\bibliography{pv}

\end{document}